\documentclass[twocolumn,showpacs,preprintnumbers,amsmath,amssymb,floatfix]{revtex4}

\usepackage{graphicx}
\usepackage{dcolumn}
\usepackage{bm}

\newcommand{\beq}{\begin{equation}}
\newcommand{\eeq}{\end{equation}}

\begin{document}

\title{An analysis on the photoproduction of massive gauge bosons at the LHeC}
\author{C. Brenner Mariotto$^{a}$ and M.V.T. Machado$^{b}$}

\affiliation{
$^a$ Instituto de Matem\'atica, Estat\'\i stica e F\'\i sica
, Universidade Federal do Rio Grande\\
Caixa Postal 474, CEP 96203-900, Rio Grande, RS, Brazil\\
$^b$ High Energy Physics Phenomenology Group, GFPAE  IF-UFRGS \\
Caixa Postal 15051, CEP 91501-970, Porto Alegre, RS, Brazil
}

\begin{abstract}
In this work we investigate the photoproduction of massive gauge bosons, $W^{\pm}$ and $Z^0$,  as part of relevant physics topics to be studied in the proposed electron-proton collider, the LHeC. The estimates for production cross sections and the number of events are presented. In addition, motivated by the intensive studies to test the deviations from the Standard Model at present and 
future colliders, we discuss the $W^{\pm}$ asymmetries and perform an analysis on the role played by anomalous $WW\gamma$ coupling. 
\end{abstract}


\maketitle

\section{Introduction}
Being planned to start around 2020/2022, the Deep Inelastic Electron-Nucleon Scattering at the LHC (LHeC) machine is a possible extension of the current LHC at CERN, an electron-proton collider \cite{dainton}. It is a convenient way to go beyond the LHC capabilities, exploiting the 7 TeV proton beams which will be produced at the LHC, to drive research on $ep$ and $eA$ physics at some stage during the LHC time. This LHC extension will open a new kinematic window - the $\gamma p$ CM energy can reach up to TeV scale, far beyond the $\sqrt{s}\gg$ 200 GeV at HERA, a very proficuous region for small-$x$ physics and many other physics studies. In particular, the energy of the incoming proton is delivered by the LHC beam, and a list of possible scenarios is considered for the energy of the incoming electron as $E_p=7$ TeV and $E_e=50-200$ GeV, corresponding to the center of mass energies of $\sqrt{s}=2\sqrt{E_pE_e}\simeq 1.18-2.37$ TeV \cite{desreport}. The anticipated integrated luminosity is of order $10-10^2$ fb$^{-1}$ that depends on the energy of the electron beam and also the machine design.

Despite of great successes of the Standard Model (SM) and difficulties in finding new physics such as supersymmetry, which is the most popular scenario, the non-abelian self-couplings of $W$, $Z$, and the photon remain poorly measured up to now. In this context, the investigation of three gauge boson couplings plays an important role to manifest the non-abelian gauge symmetry in standard electroweak theory.
Their precision measurement will be the crucial test of the structure of the
SM. The inclusive and exclusive production of $W$ and $Z$ at the LHC already provides important tests of the SM and beyond. However, the photoproduction channel has the advantage of being much cleaner than the $pp$ collision channels. The physics program of the LHeC will explore the high-energy domain complementing the LHC and its discovery potential for physics beyond the SM with the great precision DIS measurements at high luminosities. The design report already contains \cite{desreport} some estimates for a variety of electroweak interaction processes such as leptoquarks/leptogluons, new heavy leptons, new physics in boson-quark interactions, and sensitivity to a Higgs boson. In this work we investigate the photoproduction of massive gauge bosons at the TeV scale and also examine the potential of the LHeC collider to probe anomalous WW$\gamma$ coupling. Along these lines, we propose some observables that are sensitive to deviations from SM physics. Previous theoretical studies on such a subject are quite compelling, and the $WW\gamma$ vertex in $ep$ colliders was addressed in Refs. \cite{WWg1,WWg2,WWg3,WWg4} long ago.

The aim of this work is twofold - first, we show predictions for the photoproduction of massive gauge bosons at future LHeC energies within SM physics. The photoproduction cross section including the resolved and direct processes are obtained, as well as their number of events in the most promising final states decays. We then go beyond it, with the photoproduction of $W$ bosons, analyzing the production rates of $W$ bosons, as they move away from the SM. The sensitivity of the LHC for deviations from the SM is investigated, and some additional observables are proposed. This article is organized as follows. The basic formulas to calculate the photoproduction of $W^{\pm}$, $Z^0$ and virtual photons are presented in the next section, including the expressions for the anomalous coupling in the $W$-production case. Our numerical results for the photoproduction cross section and event rates within the SM and beyond are presented in section \ref{Wbeyond}, followed by the corresponding discussion. The summary and conclusions are presented in section \ref{conc}.

\section{Cross Sections in the Standard Model and beyond}

Let us start by considering the C and P parity conserving effective Lagrangian for two charged W-boson and one photon interactions \cite{hagiwara}. The motivation is to use the $W$ photoproduction cross section as a test of the $WW\gamma$ vertex. In such a case, it is introduced by two dimensionless parameters, $\kappa$ and $\lambda$, which are related to the magnetic dipole and electric quadrupole moments, namely, $\mu_W = \frac{e}{2m_W}(1+\kappa+\lambda)$ and $Q_W = -\frac{e}{m_W^2}(\kappa-\lambda)$. In the case of values $\kappa=1$ and $\lambda=0$, the SM is recovered at tree level. We are left with three diagrams for the subprocess $\gamma q_{i}\rightarrow Wq_{j}$ and only t-channel W exchange graph contributes to the $WW\gamma$ vertex. The unpolarized differential cross section for the subprocess $\gamma q_{i}\rightarrow Wq_{j} $ can be obtained using helicity amplitudes from summing over the helicities. For the signal, we are considering a quark jet and an on-shell W with leptonic decay mode $\gamma p\rightarrow W^{\mp}+jet\rightarrow\ell+p_{T}^{miss}+jet $, where $\ell=e,\mu$. In the current mode, the charged lepton and the quark jet are nicely separated and the signal is prevented from being in the background of the SM.

The cross section for the subprocess $ \gamma q_{i} \rightarrow W q_{j}$ is composed of the direct and resolved-photon production, $\hat{\sigma}=\hat{\sigma}_{dir}+\hat{\sigma}_{res}$. The direct-photon contribution is given by \cite{WWg2,WWg3}
\begin{eqnarray}
\hat{\sigma}_W &=&\sigma_0\{|V_{q_{i}q_{j}}|^{2}\{(|e_{q}|-1)^{2}(1-2\hat{z}+2\hat{z}^{2})
\log({\hat{s}-M_{W}^{2}\over\Lambda^{2}}) \nonumber \\
& - & [(1-2\hat{z}+2\hat{z}^{2}) - 2|e_{q}|(1+\kappa+2\hat{z}^{2})
+{{(1-\kappa)^{2}}\over{4\hat{z}}} \nonumber \\
&- & {{(1+\kappa)^{2}}\over{4}}]
\log{\hat{z}}+  [(2\kappa+{{(1-\kappa)^{2}}\over{16}})
{1\over \hat{z}} \nonumber \\
& +& ({1\over 2}
 +  {{3(1+|e_{q}|^{2})}\over{2}})\hat{z}
 + (1+\kappa)|e_{q}|-{{(1-\kappa)^{2}}\over{16}} \nonumber \\
& + & {|e_{q}|^{2}\over 2}](1-\hat{z})
-{{\lambda^{2}}\over{4\hat{z}^{2}}}(\hat{z}^{2}-2\hat{z}
\log{\hat{z}}-1)\nonumber \\
&+ &{{\lambda}\over{16\hat{z}}}
(2\kappa+\lambda-2)[(\hat{z}-1)(\hat{z}-9)
+4(\hat{z}+1)\log{\hat{z}}]\}, \nonumber \\
\label{sigmadir}
\end{eqnarray}
where  $\sigma_0 ={{\alpha G_{F}M_{W}^{2}}\over{\sqrt{2}\hat{s}}}$, $\hat{z}=M_{W}^{2}/\hat{s}$ and $\Lambda^{2}$ is the cutoff scale in order to regularize the $\hat{u}$-pole of the
collinear singularity for massless quarks. In addition, $\Lambda^2$ is the scale that determines the running of photon structure functions in the resolved part. The quantity $V_{ij}$ is the Cabibbo-Kobayashi-Maskawa (CKM) matrix and  $e_{q}$ is the quark charge.

The direct part of the cross section then reads
\begin{eqnarray}
\sigma_{dir}(\gamma p \rightarrow W^{\pm}X)=\int_{x_p^{m}}^{1}dx_p\sum_{q,\bar{q}}f_{q/p}(x_p,Q^{2})\,\hat{\sigma}_W(\hat{s}),
\end{eqnarray}
where $f_{q/p}$ are the parton distributions functions in the proton, $x_p^{m}=M_W^2/s$ and $\hat{s}=x_ps$.

The resolved-photon part of the cross section can be calculated 
using the usual electroweak formula for the $q_{\gamma}q_{p}\to W^{\pm}$ fusion process, $\hat{\sigma}(q_i\bar{q}_j\rightarrow W)=\frac{\sqrt{2}\pi}{3}G_Fm_W^2|V_{ij}|^2\delta (x_ix_js_{\gamma p}-m_W^2)$. For the photoproduction cross sections one needs parton distribution
functions inside the photon and proton. The photon structure
function $f_{q/\gamma}$ consists of perturbative pointlike parts 
and hadronlike parts. Putting it all together, the resolved-photon part reads
\begin{eqnarray}
& & \sigma_{res}(\gamma p \rightarrow W^{\pm}X)  =  \frac{\pi\sqrt{2}}{3\,s}G_{F}m_{W}^{2}
|V_{ij}|^{2}\int_{x_{\gamma}^m}^{1}\frac{dx_{\gamma}}{x_{\gamma}}\nonumber \\
&\times & \sum_{q_i,q_j}f_{q_{i}/p}
(\frac{m_{W}^{2}}{xs},Q_{p}) \left[f_{q_{j}/\gamma}(x_{\gamma},
Q_{\gamma}^{2})-\tilde{f}_{q_{j}/\gamma}(x_{\gamma},Q_{\gamma}^{2})\right], \nonumber \\
\end{eqnarray}
where in order to avoid double counting on the leading
logarithmic level, one subtracts the pointlike part of the photon structure function (photon splitting at large $x$), $\tilde{f}_{q/\gamma}(x,Q_{\gamma}^{2})=\frac{3\alpha e_{q}^{2}}{2\pi}[x^{2}+(1-x)^{2}]\log (Q_{\gamma}^{2}/\Lambda^{2})$. In addition, here $x_{\gamma}^m=M_W^2/s$.

Similar calculation can be done for the $Z$ boson photoproduction. Here, we focus on the SM prediction. Once again, the cross section for the subprocess $ \gamma q \rightarrow Z q$ is composed of the direct and resolved-photon production, $\hat{\sigma}=\hat{\sigma}_{dir}+\hat{\sigma}_{res}$. The direct-photon contribution is given by
\begin{eqnarray}
\hat{\sigma}_Z & = &  \frac{\alpha G_{F}M_{Z}^{2}}{\sqrt{2}\,\hat{s}}\,
g_q^2e_q^2\, \left[ \left(1-2\hat{z}+2\hat{z}^{2}\right)\log \left(\frac{\hat{s}-M_{Z}^{2}}{\Lambda^{2}}\right) \right. \nonumber \\
& + & \left. \frac{1}{2}\left(1+2\hat{z}-3\hat{z}^{2}\right)\right],
\end{eqnarray}
where now $\hat{z}=M_{Z}^{2}/\hat{s}$ and $g_q^{2}= \frac{1}{2}(1-4|e_e|x_W+8e_q^2x_W^2)$, with $x_W = 0.23$. 

The direct part of the $Z$-photoproduction cross section then reads
\begin{eqnarray}
\sigma_{dir}(\gamma p \rightarrow Z^0X)=\int_{x_p^{m}}^{1}dx_p\sum_{q,\bar{q}}f_{q/p}(x_p,Q^{2})\,\hat{\sigma}_Z(\hat{s}),
\end{eqnarray}
where $f_{q/p}$ are the parton distributions functions in the proton, $x_p^{m}=m_Z^2/s$. 

The resolved-photon part of the cross section stands for the subprocess $q\bar{q}\to Z^0$, and it is written as
\begin{eqnarray}
& & \sigma_{res}(\gamma p \rightarrow Z^0X)  =  \frac{\pi\sqrt{2}}{3\,s}G_{F}m_{W}^{2}g_q^2\int_{x_{\gamma}^m}^{1}\frac{dx_{\gamma}}{x_{\gamma}}\nonumber \\
&\times & \sum_{q}f_{\bar{q}/p}
\left(\frac{m_{Z}^{2}}{xs},Q_{p}\right) \left[f_{q/\gamma}(x_{\gamma},
Q_{\gamma}^{2})-\tilde{f}_{q/\gamma}(x_{\gamma},Q_{\gamma}^{2})\right]. \nonumber \\
\end{eqnarray}

In the next section we compute the numerical results for the $W^{\pm}$ and $Z^0$ photoproduction cross section in the LHeC regime of energy/luminosity. We also investigate the sensitivity to anomalous $WW\gamma$ couplings associated to beyond SM physics.

\section{Numerical results and discussions}
\label{Wbeyond}

\begin{figure}[t]
\includegraphics[scale=0.4]{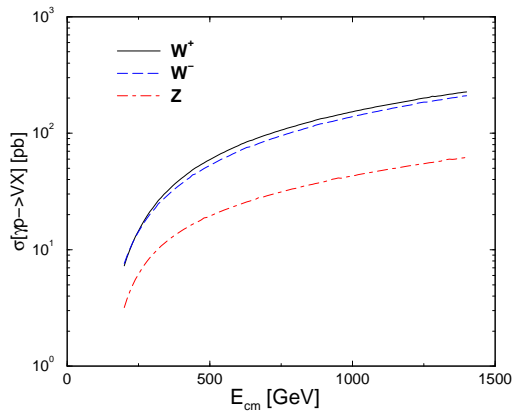}
\caption{Cross sections for the production of massive $W^{\pm}$ and $Z^0$ gauge bosons as a function of the CM energy.}
\label{fig:1}
\end{figure}

Let us now perform a preliminary study for the LHeC machine \cite{dainton,desreport}. Using the design with an electron beam having laboratory energy of $E_e=70$ GeV, the center of mass energy will reach $E_{cm}=W_{\gamma p}=1.4$ TeV and a nominal luminosity of order $10^{33}$ cm$^{-2}$s$^{-1}$. Our estimates for the massive boson photoproduction cross sections in the SM are the following. One gets $\sigma(\gamma +p\rightarrow W^{\pm}X)\simeq 400$ pb and $\sigma(\gamma +p\rightarrow Z^0X)\simeq 60$ pb. These are roughly estimates, since we have not introduced the $K$-factors associated with next-to-leading-order (NLO) corrections to the processes. We have summed the resolved and direct contributions. The energy behavior for the cross section is presented in Fig. \ref{fig:1}. The dependence is quantitatively given by $\sigma_V\propto W_{\gamma p}^{\alpha}$, with $\alpha \simeq 1.312$. It is seen that the cross sections are at least one order of magnitude larger than for DESY-HERA machine, $W_{\gamma p }\simeq 300$ GeV.

In Table \ref{tab1} the photon-proton total cross sections times branching
ratio of $W\rightarrow \mu\nu $ and the corresponding number of events
 are shown for SM parameters for W ($\kappa=1$ and $\lambda=0$) and also for the $Z^0$ boson with a corresponding branching
ratio of $Z^0\rightarrow \mu^+\mu^- $. The number of events has been computed using $N_{ev}=\sigma(e p\rightarrow V+X)BR(V\rightarrow \mu\nu /\mu^+\mu^-)L_{int}$. At this point we consider the acceptance in the leptonic channel as 100\%. The photoproduction cross section is calculated by convoluting the Weizs\"{a}cker-Williams spectrum
\begin{eqnarray}
f_{\gamma/e}(y)& = & \frac{\alpha}{2\pi}\left[\frac{1+(1-y)^2}{y}\log \frac{Q^2_{max}}{Q^2_{min}} \right. \nonumber \\
& - & \left.2m_e^2y\,\left(\frac{1}{Q^2_{min}}- \frac{1}{Q^2_{max}}\right)  \right],
\end{eqnarray}
with the differential hadronic cross section. Here, $Q^2_{min}=m_e^2y/(1-y)$ and we impose a cut of $Q^2_{min}=0.01$.

Through the calculations, proton
structure functions of CTEQ \cite{Pumplin:2002vw} and photon structure functions
of GRV \cite{grvphoton} have been used
with $Q^{2}=M_{W}^{2}$. The usual electroweak parameters are taken from Ref. \cite{PDG}. We have assumed an integrated
luminosity $L_{int}$ at 10 fb$^{-1}$ \cite{desreport} in order to compute the number of events, $N_{ev}$. The number of events is large enough to put forward further analysis, as we have units of events per second for $W^{\pm}$.

\begin{table}[ht]
\caption{The photon-proton cross sections times branching ratios
$\sigma(\gamma p\rightarrow W^{\pm}X)\times BR(W^{+}\rightarrow
 \mu\nu)$  and $\sigma(\gamma p\rightarrow Z^0X)\times BR(Z^0\rightarrow
 \mu^+\mu^-)$ in units of pb. The number of events $N_{ev}$ is also presented at an integrated luminosity 10 fb$^{-1}$.
\label{tab1}}
\begin{center}
\begin{tabular}{|l|c|c|}
\hline
  $V$  & $\sigma(\gamma p\rightarrow V\,X) \times BR$& $N_{ev}$ \\
\hline
$W^+$ & 24 & $1.2\times 10^{4}$ \\
$W^-$ & 24 & $1.2\times 10^{4}$ \\
$Z^0$ & 2.1 & $1.1\times 10^{3}$ \\
\hline
\end{tabular}
\end{center}
\end{table}

Let us now investigate the scenario for physics beyond the SM. Certain properties of the $W$ bosons, such as the magnetic dipole and the electric quadrupole moment, play a role in the interaction vertex $WW\gamma$, thus processes involving this vertex offer the opportunity to measure such properties. The magnetic dipole moment $\mu_W$
and the electric quadrupole moment $Q_W$ of the $W$ bosons can be written in terms of parameters $\kappa, \,\lambda$, where $\kappa=1$ and $\lambda=0$ are the Standard Model values for those parameters at tree level. According to the Particle Data Group \cite{PDG}, the measured value of $\mu_W/\frac{e}{2M_W}=1+\kappa + \lambda=2.22\pm 0.20$ suggests that there are deviations from the standard values. In $W$ photoproduction one has a unique scenario to test the anomalous $WW\gamma$ vertex and its $\kappa$ and $\lambda$ parameters. The $WW\gamma$ vertex [$W^+(p_1)$, $W^-(p_2)$, $A(p_3)$], denoted by $\Gamma_{\mu\nu\rho}(p_1,p_2,p_3)$,  is given by \cite{Dubinin}
\begin{eqnarray}
\frac{\Gamma_{\mu\nu\rho}}{e} & = & \left[ g_{\mu\nu}\left(p_1-p_2-\frac{\lambda}{M_W^2}[(p_2\cdot p_3)p_1-(p_1\cdot p_3)p_2]\right)_{\rho}
\right.\nonumber\\
& + & g_{\mu\rho}\left({\kappa} p_3-p_1+\frac{\lambda}{M_W^2}[(p_2\cdot p_3)p_1-(p_1\cdot p_2)p_3]\right)_{\nu}  \nonumber\\
&+ & g_{\nu\rho}\left(p_2-{\kappa} p_3-\frac{\lambda}{M_W^2}[(p_1\cdot p_3)p_2-(p_1\cdot p_2)p_3]\right)_{\mu} \nonumber \\
& + & \left. \frac{\lambda}{M_W^2}\left(p_{2\mu}p_{3\nu}p_{1\rho}-p_{3\mu}p_{1\nu}p_{2\rho}\right) \right]
\end{eqnarray}
where the anomalous contributions from the SM are taken into account if they are included in the terms involving $\kappa\ne 1$ and/or $\lambda\ne 0$.

In the photoproduction of $W^{\pm}$ bosons, the direct contribution $\sigma_{dir}$ involves the generalized $WW\gamma$ vertex, and then the expression for $\hat{\sigma}_W(\hat{s}=x_ps)$ in Eq. (\ref{sigmadir}) can be used to investigate deviations from SM physics. An interesting observable is the number of muon plus neutrino events coming from the decay of the $W^+$. This is shown in Table \ref{tab2}, where we assumed  the luminosity of ${\cal {L}}=10\,$ fb$^{-1}$. As we can see, the number of $W^+\to\mu\nu$ events is very dependent on the choice of the $\kappa$ and $\lambda$ parameters, and in most scenarios it increases as $\kappa$, $\lambda$ increase/depart from Standard Model. Such an effect could certainly be tested at LHeC.
\begin{figure*}[ht]
\begin{center}
\includegraphics[scale=0.4]{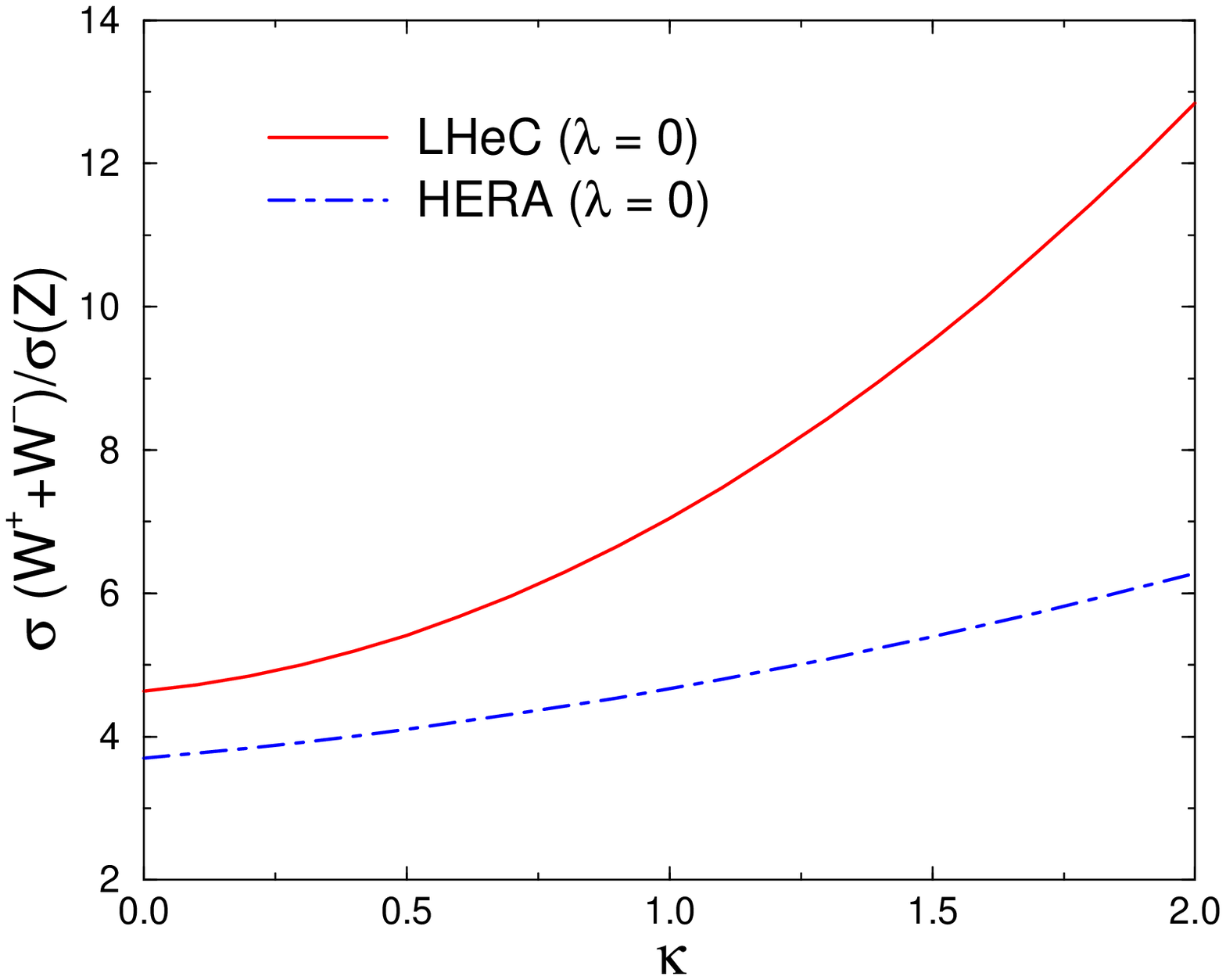} 
\includegraphics[scale=0.4]{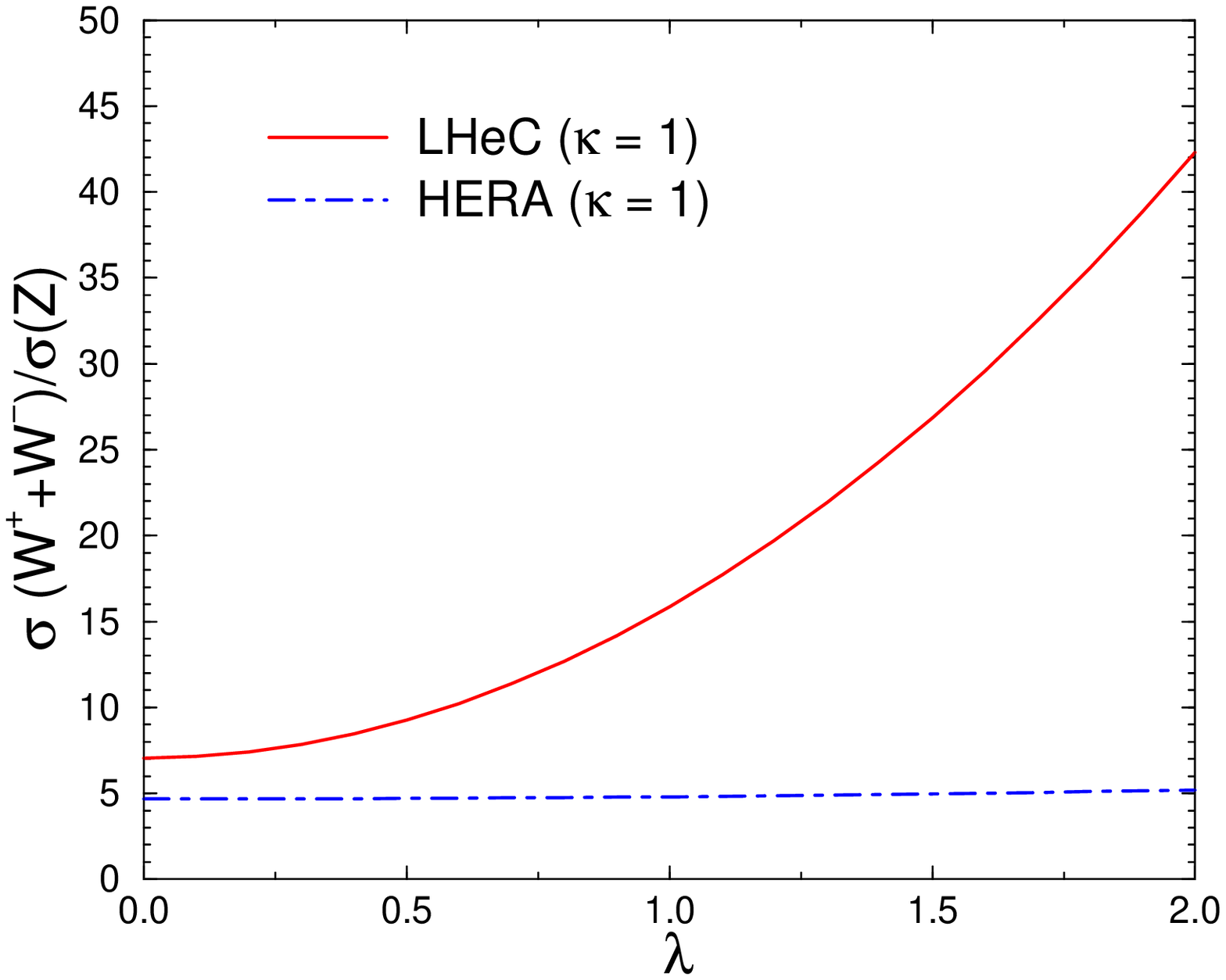}
\caption{The ratio $R_{W/Z}(\kappa,\lambda)$ for both DESY-HERA and LHeC energies. Left panel: ratio as a function of the $\kappa$ parameter for fixed $\lambda =0$ (SM value). Right panel: ratio as a function of the $\lambda$ parameter for fixed $\kappa =1$ (SM value).}
\label{ratiowz}
\end{center}
\end{figure*}

\begin{table}[ht]
\caption{The number of muon plus neutrino events coming from the $W^+$ decay for distinct choices for the parameters $\kappa$ and $\lambda$ presented at an integrated luminosity of 10 fb$^{-1}$.
\label{tab2}}
\begin{center}
\begin{tabular}{|l|l|c|c|}
\hline
  $\kappa$ & $\lambda$  & $\sigma(\gamma p\rightarrow W^+\,X) \times BR$ [pb] & $N_{ev}$ \\
\hline
 0 & 0 & 16 & $8\times 10^{3}$ \\
 1 & 0 & 24 & $1.2\times 10^{4}$ \\
 2 & 0 & 44 & $2.2\times 10^{4}$ \\
 1 & 1 & 61 & $3.1\times 10^{4}$ \\
 1 & 2 & 172 & $8.5\times 10^{4}$\\
\hline
\end{tabular}
\end{center}
\end{table}

As we have already noticed, the cross sections for the $W$ and $Z$ production may have contributions due to higher-order terms that where not included in our calculation. Cross sections in NLO considering the default Standard Model vertices were already calculated in \cite{nlospira}. In order to get rid of normalization uncertainties, one can take the ratios $\sigma_W^{\pm}/\sigma_Z$ to test the $WW\gamma$ vertex and the $\kappa$ and $\lambda$ parameters. To do this we propose the study of the following observable:
\begin{eqnarray}
R_{W/Z} (\kappa,\lambda;\sqrt{s}) = \frac{\sigma_{W^+} +\sigma_{W^-}}{\sigma_Z}\,,
\end{eqnarray}
which can be constructed from equations given in the previous section. Such an observable was already proposed some time ago, in Refs. \cite{WWg2,WWg3}.  In Fig. \ref{ratiowz} we show our results for the $R_{W/Z}$ ratio, for HERA and LHeC energies. In the left plot, the dependence of the ratio is shown as a function of the $\kappa$ parameter for fixed $\lambda=0$. In addition, in the right plot the ratio is presented as a function of the $\lambda$ parameter for fixed $\kappa = 1$. The results are presented for both DESY-HERA and LHeC energies. We can study the sensitivity with $\kappa$ and $\lambda$ parameters, regarding the SM and possible new physics.  The results show that the ratio has much more sensitivity to the $\kappa$ and $\lambda$ parameters for LHeC energies. Regarding the $\lambda$ parameter, it is unimportant at HERA energies. Thus, the LHeC collider would be able to pin down the correct values for these parameters and then determine the magnet dipole and electric quadrupole of the $W$.

\begin{figure*}[ht]
\begin{center}
\includegraphics[scale=0.4]{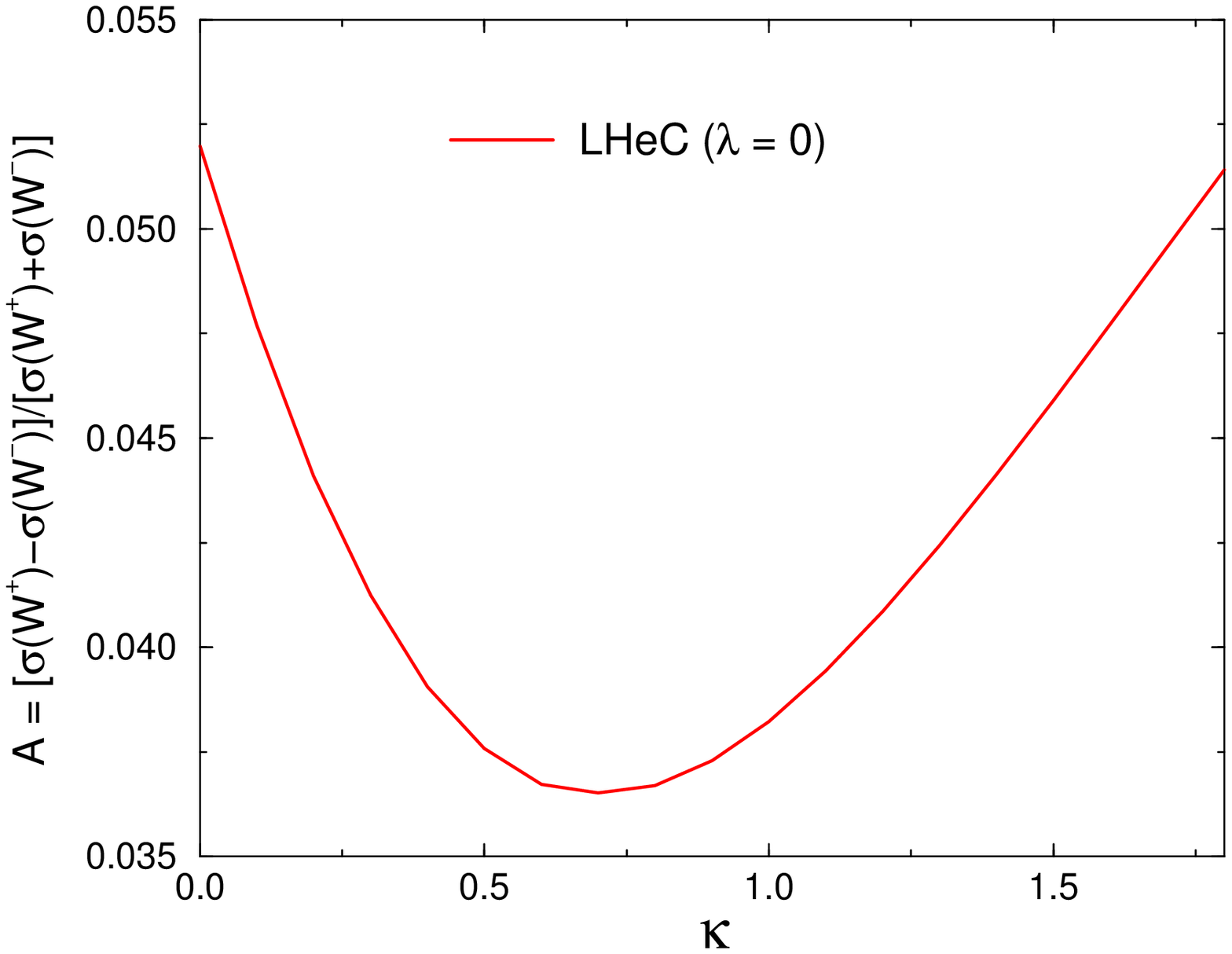}
\includegraphics[scale=0.4]{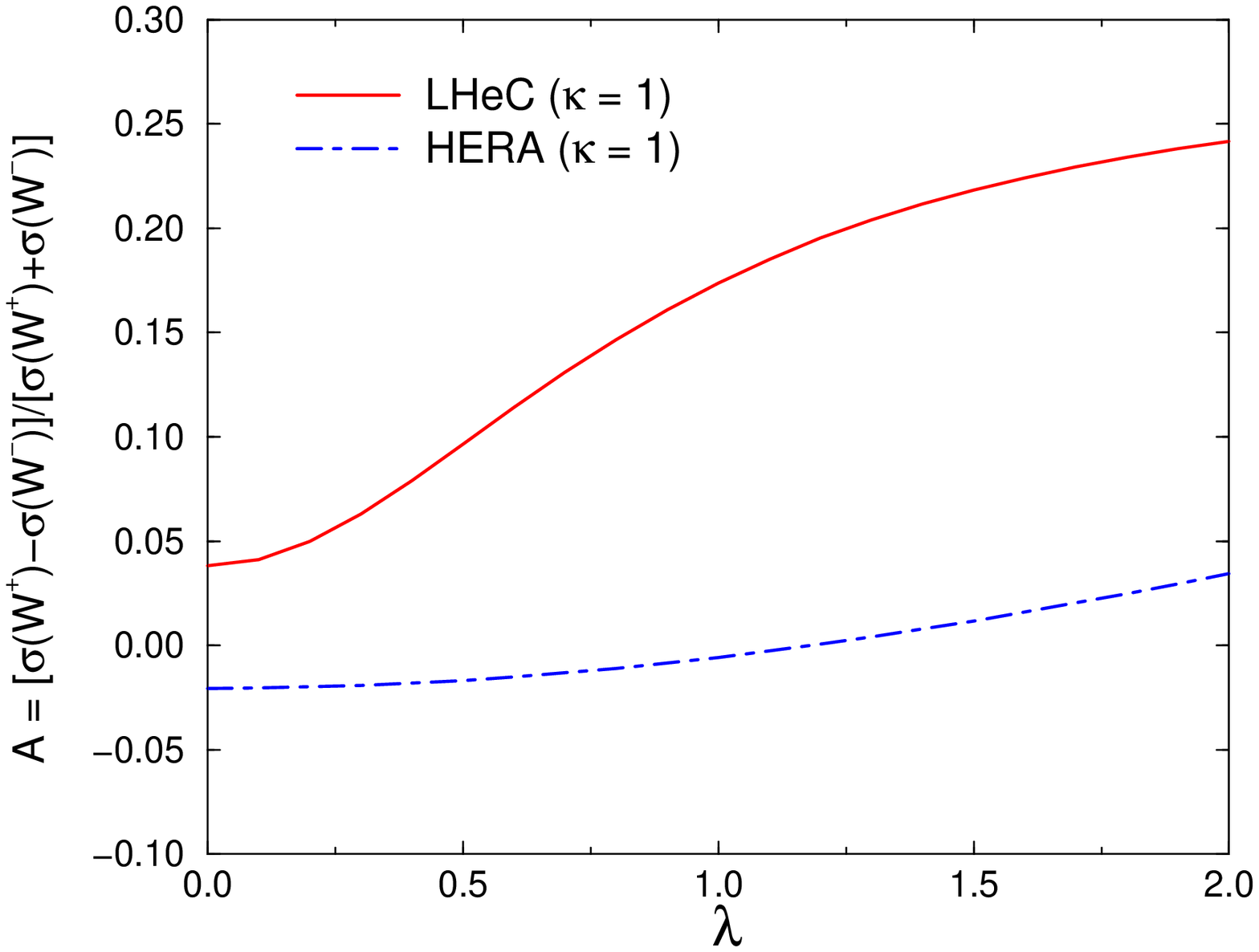}
\caption{The asymmetry $A(\kappa,\lambda;\sqrt{s})$ at the LHeC regime. Left panel: asymmetry as a function of the $\kappa$ parameter for fixed $\lambda =0$ (SM value). Right panel: asymmetry as a function of the $\lambda$ parameter for fixed $\kappa =1$ (SM value). In this last case, the result for DESY-HERA energy is also shown.}
\label{asy}
\end{center}
\end{figure*}

Another observable that could be studied and tested at the LHeC is the $W^{+}W^{-}$ asymmetry, defined by

\begin{eqnarray}
A(\kappa,\lambda;\sqrt{s})  = \frac{\left(\sigma_{W^+} - \sigma_{W^-}\right)}{\left(\sigma_{W^+} +\sigma_{W^-}\right)}.
\end{eqnarray}

The results for this asymmetry are shown in Fig. \ref{asy}, where we show the sensitivity with the $\kappa$ and $\lambda$ parameters. As we can see, for LHeC energies the $W^{+}W^{-}$ asymmetry depends strongly on the $\kappa$ and $\lambda$ parameters and is therefore a useful observable to help determine the best scenarios. In the left plot we show the dependence of asymmetry as a function of the $\kappa$ parameter for fixed $\lambda=0$ (its SM value) only for the LHeC energy. Moreover, in the right plot we present the ratio as a function of the $\lambda$ parameter for fixed $\kappa = 1$, where the corresponding result for DESY-HERA energy is also shown. As a general conclusion about the anomalous coupling, we see that with the LHeC collider the parameters $\kappa$ and $\lambda$ have better sensitivity than DESY-HERA $ep$ collider, and it would give complementary information to the LHC collider.

Finally, let us compare the present calculation to previous studies. In Ref. \cite{WWg3} the massive boson photoproduction is considered for an energy of $\sqrt{s}\simeq 1.3$ TeV and integrated luminosity of 1 fb$^{-1}$. For the SM values of $\kappa$ and $\lambda$ parameters, the photoproduction cross section was obtained as $11.3$ pb, $12.2$ pb and $5.4$ pb for $W^+$, $W^-$ and $Z^0$, respectively. The number of events for $Z^0$ was estimated to be 360. These results are completely consistent with ours when considering the integrated luminosity of 10 fb$^{-1}$. Concerning the sensitivity to the parameters, we found the general trend is similar; however, the cross sections for higher values of parameters are relatively larger than ours. We have checked that our ratio $\sigma(W^{\pm})/\sigma(Z)$ is 50 \% smaller than in Ref. \cite{WWg3} for several values of parameters $\kappa,\,\lambda$. This probably is due the different energy and the theoretical uncertainties coming from the PDFs considered.

In Ref. \cite{WWg4} an analysis quite similar to ours was performed focusing on the spectrum of the backscattered laser photon (energy of $\sqrt{s}=1.7$ TeV and integrated luminosity of 200 pb$^{-1}$). For a Weizs\"{a}cker-Williams spectrum, we have checked that the numbers of events in process $\gamma p \rightarrow W^+\mathrm{jet}$ is quite consistent with ours when considering the same integrated luminosity. Their original values are, for instance, 288 and 1151 events for sets ($\kappa =1,\,\lambda=0$) and ($\kappa=1,\,\lambda=2$), respectively. The compatibility is good, as we are using a luminosity 100 times larger.

The photoproduction of the $W$ boson at HERA has been addressed at the NLO level of QCD corrections in Ref. \cite{nlospira}. The prediction is $\sigma(W^+)=0.478$ pb and $\sigma(W^-)=0.484$ pb at $\sqrt{s}=318$ GeV, imposing a cut $p_T<25$ GeV. The main conclusion is that the QCD corrections reduce the factorization scale dependence significantly and modify the leading-order prediction by a factor 10\%. This can be used as a good argument for our LO calculations here. In addition, we can rescale their prediction for the LHeC case. A rough estimate would give $\sigma(W^+)\times BR=0.33$ pb and $\sigma(W^-)\times BR=0.34$ pb at the LHeC. This is smaller than our results in Table 1, where one has $\sigma(e+p\rightarrow W^{\pm}+X)\times BR=1.2$ pb, which is is associated with the cut on boson transverse momentum and distinct kinematic cuts in the integration of the Weizs\"{a}cker-Williams spectrum.

\section{Summary}
\label{conc}

We have examined the prospects for massive gauge bosons detection at the proposed Deep Inelastic Electron-Nucleon Scattering at the LHC (LHeC) machine. The photon-proton cross sections have been computed for $W^{\pm}$ and $Z^0$ inclusive production and are of the order of dozens of picobarns. The number of events is evaluated for the photoproduction cross section, assuming an integrated luminosity of 10 fb$^{-1}$ and, they are large enough to make the measurements feasible. We have also investigated the anomalous $WW\gamma$ coupling using the machine design. We found that the likelihood of that kinematic limit to be available at the LHeC is somewhat increased relative to the previous DESY-HERA machine. We have tested some sample scenarios beyond SM physics by scanning the values of parameters $\kappa$ and $\lambda$ considering anomalous $WW\gamma$ coupling. In the case of anomalous coupling, the photoproduction process at the LHeC proves to be a powerful tool. Finally, we consider different observables that together could contribute to pinning down the correct $WW\gamma$ vertex. We introduced the ratio $\sigma (W^{\pm})/\sigma (Z)$, which is less sensitive to the NLO QCD corrections and the $W$-asymmetry observable $A(\kappa,\lambda;\sqrt{s})$ that scanns asymmetries in the $W$-photoproduction.

\begin{acknowledgments}
 This research was supported by CNPq and FAPERGS, Brazil. 
\end{acknowledgments}

\end{document}